\documentclass[pdflatex,sn-mathphys-num]{sn-jnl}

\usepackage{graphicx}%
\usepackage{multirow}%
\usepackage{amsmath,amssymb,amsfonts}%
\usepackage{amsthm}%
\usepackage{mathrsfs}%
\usepackage[title]{appendix}%
\usepackage{xcolor}%
\usepackage{textcomp}%
\usepackage{manyfoot}%
\usepackage{booktabs}%
\usepackage{algorithm}%
\usepackage{algorithmicx}%
\usepackage{algpseudocode}%
\usepackage{listings}%
\usepackage{ulem}

\theoremstyle{thmstyleone}%
\theoremstyle{thmstyletwo}%

\theoremstyle{thmstylethree}%

\begin{document}

\title[Article Title]{MHz to sub-kHz field detection with an all-dielectric potassium Rydberg-atom sensor}


\author[1,3]{\fnm{Daniel} \sur{Hammerland}} 
\author[1,3]{\fnm{Rajavardhan} \sur{Talashila}}
\author[1,3]{\fnm{Dixith} \sur{Manchaiah}}
\author[3]{\fnm{Noah} \sur{Schlossberger}}
\author[3]{\fnm{Nikunjkumar} \sur{Prajapati}}
\author[2]{Erik~McKee}
\author[2]{Michael~A.~Highman}
\author[3]{\fnm{Matthew T.} \sur{Simons}}
\author[3]{\fnm{Samuel} \sur{Berweger}}
\author[3]{\fnm{Alexandra B.} \sur{Artusio-Glimpse}}
\author*[3]{\fnm{Christopher L.} \sur{Holloway}}\email{christopher.holloway@nist.gov}

\affil[1]{\orgdiv{Department of Physics}, \orgname{University of Colorado, Boulder}, \orgaddress{\street{Libby Dr}, \city{Boulder}, \postcode{80305}, \state{CO}, \country{USA}}}
\affil[2]{TOPTICA Photonics, Pittsford, New York 14534, USA}
\affil[3]{\orgdiv{Communications Technology Division}, \orgname{National Institute of Standards and Technology}, \orgaddress{\street{325 Broadway}, \city{Boulder}, \postcode{80305}, \state{CO}, \country{USA}}}


\abstract{Rydberg sensors have significant promise as an alternative to the antenna systems used for sub-MHz frequency communications, where the scale of high-efficiency antennas is often impractically large, forcing the use of low-efficiency, electrically small antennas. The exploration of Rydberg sensors at these frequencies has been hampered by the low field transmission of the silicate vapor cells. We dramatically improve the low-frequency field transmission of silicate vapor cells by using potassium as the active medium instead of rubidium or cesium. The potassium Rydberg sensor can measure fields with frequencies down to 500~Hz in an all-dielectric sensor, effectively extending the low-frequency cutoff of the sensor by nearly four orders of magnitude compared to an equivalent rubidium vapor cell. With this simple substitution, experimentation with low-frequency sensing becomes dramatically more accessible to the community. 
}


\maketitle

\section{Introduction}\label{sec1}
In classical receiver systems, antennas are used to detect radio frequency (RF) signals. An antenna is most efficient when the antenna size is on the order of half the wavelength of the signal frequency, $\frac{\lambda}{2}$\cite{balanis2016antenna}. For a 1 MHz signal, this places the antenna on the scale of a hundred meters, and at 1 kHz, that size extends to hundreds of kilometers. In many applications, antennas at these sizes are prohibitively large, which forces the use of electrically-small antennas\cite{fano_theoretical_1947, hansen_fundamental_1981}. Electrically small antennas suffer from low efficiency and bandwidth. While matching circuits can be used to improve their efficiency, they reduce the bandwidth that the antenna can receive. 


Rydberg sensors~\cite{6910267, schlossberger_rydberg_2024} operate on fundamentally different principles. In these systems, the energy levels of highly polarizable Rydberg states of atoms are perturbed by incident RF fields. The perturbation of the Rydberg states is imprinted onto an optical transmission spectrum, providing an optical readout for the RF field. As the interactions between the atom and the electric field are conservative in nature~\cite{grimm_optical_2000}, power transfer to the atomic system is not necessary for signal reception. This liberates Rydberg sensors from the scaling laws and impedance matching challenges of classical antennas, making them a highly attractive addition to modern communication systems. Despite this promise, sensing low-frequency electric fields has been an ongoing challenge for Rydberg field sensors, even at~MHz ranges~\cite{liu_highly_2022, rotunno_detection_2023}.

In Rydberg sensors, an atomic species is placed in glass vapor cells. To date, rubidium and cesium have been exclusively used, mainly because of their high vapor pressure and the commercial availability of lasers needed to excite electrons into Rydberg states. Although glass is an insulator, the addition of rubidium and cesium results in a dramatic screening at frequencies below a few~MHz~\cite{bouchiat_electrical_1999, richardson_extraction_2025,Ma:20}. There are several pathways that the field has explored to develop low-frequency Rydberg sensors. Plated vapor cells, where a set of metallic electrodes is placed inside the walls of the vapor cell, have been able to measure fields down to DC~\cite{holloway_electromagnetically_2022, ma_measurement_2022, li_super_2023,chen_atomic_2024,10.1116/5.0090892}. However, these cells require electrical contact to operate well, which restricts them from being used in any application with free-space fields. Sapphire vapor cells demonstrate excellent low-frequency field transmission~\cite{bouchiat_electrical_1999, richardson_extraction_2025} and offered a modified EIT signal to sub-kHz fields~\cite{jau_vapor-cell-based_2020}, although no sensitivity measurements are reported. However, the benefits of sapphire are accompanied by several substantial drawbacks. For electric fields, the high refractive index reduces RF field transmission at nearly all RF frequencies, and birefringence can distort optical polarizations. Of greater consequence than dielectric losses are the manufacturing challenges associated with sapphire. In contrast to glasses, which have a ``softened" phase, sapphire transitions from a solid to a liquid, preventing it from being blown with standard glass blowing techniques. Diffusion bonded vapor cells ~\cite{sarkisyan_alkali-vapor_2005, sekiguchi_spectroscopic_2018, jau_vapor-cell-based_2020} have removed the need for chemically reactive adhesives, but are expensive and not widely available and, consequently, have not seen widespread use. While these developments are of indisputable value, developing an all-dielectric, broadband Rydberg sensor out of materials that facilitate large-scale production would be a substantial step towards the real-world deployment of these next generation low-frequency quantum sensors. 

Here, we use potassium as the atomic species and demonstrate a substantial improvement in low-frequency field performance in an all-dielectric fused silica vapor cell, measuring fields at frequencies down to 500~Hz. To establish a benchmark for these potassium results, we repeat the measurements with a rubidium vapor cell of the same material. We find, under virtually identical conditions, that the rubidium sensor experiences substantially reduced low-frequency field transmission, resulting in dramatically worse sensitivities relative to potassium. The sensitivity measurements are further corroborated by comparing the Stark shifts at different frequencies between the two species, where, for equivalent fields and frequencies, potassium undergoes a substantially larger shift. These developments provide a highly accessible way for low-frequency sensing to be explored. In addition, they highlight the importance of understanding the chemistry between alkali metals and their vapor cells in the context of Rydberg sensors.  

\section{Results}\label{sec:Results}
\subsection{Inverted ladder scheme EIT in potassium}
\begin{figure}
    \centering
    \includegraphics[width=0.5\linewidth]{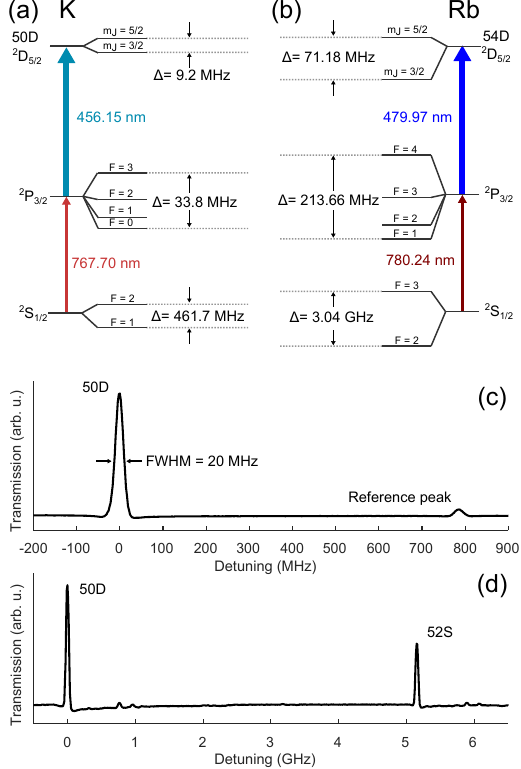}
    \caption{Potassium EIT optical transmission spectra covering (a) the 50D state and (b) the 50D and 52S states .}
    \label{fig:EIT}
\end{figure}

The results shown in this paper utilize an inverted ladder scheme of EIT that has not been explored in the literature. This has important implications for the sensitivity measurements, so  we briefly discuss the EIT spectrum of potassium. The energy level structure of potassium, Figure~\ref{fig:exp}(a), presents several notable distinctions from that of rubidium, Figure~\ref{fig:exp}(b). The energy gaps between the hyperfine states of potassium at all relevant states are roughly an factor of 14 smaller than those in rubidium. In rubidium EIT experiments, the separation between the m$_\textrm{J}$ Rydberg states can be used to easily calibrate the energy axis during a measurement.  In potassium, the separation of the 50D$_{5/2}$ state and the 50D$_{3/2}$ state is approximately 9~MHz. This becomes challenging as potassium's lighter mass and higher probe frequency increase the Doppler linewidth by nearly a factor of 1.5 compared to rubidium. This dominates over the slightly improved Doppler cancellation of the excitation beams, resulting in a broader linewidth compared to that of rubidium, for equivalent temperatures and Rabi rates. These factors, in addition to power broadening, result in an EIT linewidth of 20~MHz, Figure~\ref{fig:EIT}(c). This large width obscures the D$_{5/2}$ separation from the D$_{3/2}$, preventing these states from being used for energy calibration. Furthermore, the different F states of the 3P of potassium are much more closely packed than in rubidium. After the Doppler correction, the 33.8 MHz compresses down to 23 MHz, which means that the potassium EIT signal in Figure~\ref{fig:EIT}(c) also contains the 3P hyperfine states as well. 
The next identifiable state from the 50D peak is the 52S Rydberg state, approximately 5.1~GHz away, Figure~\ref{fig:EIT}(d). This is used to reference a weak peak that is 768~MHz away from the 50D peak. This small peaks shifts in parallel with the 50D peak during Stark shift measurements, preventing it from being used for frequency axis calibration during Stark shift measurements. This leaves the 52S state as the only option for calibration. The 5.1~GHz separation between the two peaks means that small changes in energy are not detectable. For frequencies below 1~MHz, where field transmission is low, the prominent behavior of the applied field is a small alteration of the peak height of the EIT, compared to a more discernible frequency shift. As a result, we are not confident in our ability to calibrate the ratio of the internal to external field strength below 1~MHz. 

\subsection{Comparison between potassium and rubidium}
To substantiate the claim that this emergent low-frequency field transmission originates in the substitution of a heavier alkali metal with potassium, we performed Stark shifting measurements and heterodyne sensitivity measurements in both potassium and rubidium. Several steps are taken to make the measurements as comparable as possible. Briefly, measurements in both experiments used the same optical beam paths, detectors, and RF sources. Both experiments used fused silica vapor cells, and they shared the same heating elements and thermal insulation. The n~=~50 state for potassium and the n~=~54 state of rubidium are selected because they have similar DC polarizabilities. Optical powers are adjusted to maintain Rabi rates between measurements. Further details of the setup can found in Section~\ref{se:Methods}.

\begin{figure}
    \centering
    \includegraphics[width=0.5\linewidth]{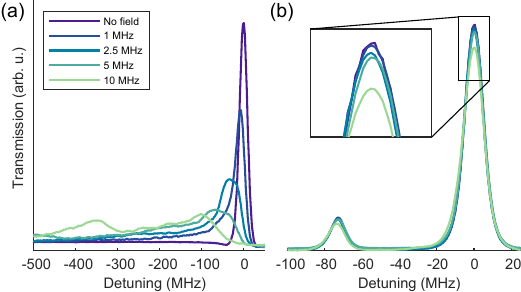}
    \caption{Comparison of the Stark shifted EIT observed in potassium \textbf{(a)} and rubidium \textbf{(b)} at different frequencies with an external field amplitude of 60 V/m.}
    \label{fig:StarkShift}
\end{figure}

Stark shift measurements in potassium and rubidium highlight the differences in the transmission of low frequency fields. Figure~\ref{fig:StarkShift} shows the measured Stark shifted spectra in the 50D state of potassium, (a), and the 54D state of rubidium, (b), at an external field strength of 60~V/m for frequencies between 1~MHz and 10~MHz. At 1~MHz, the potassium spectrum shows a 40\% reduction in the amplitude of the EIT peak compared to the measurements without an applied field. At higher frequencies, the large frequency shifts over several hundred~MHz indicate a substantial field transmission. However, rubidium measurements do not show any variation at 1~MHz, as shown in the inset of Figure~\ref{fig:StarkShift}(b).At higher frequencies, rubidium shows only a small alteration in peak height. These simple measurements emphasize the improved low-frequency field transmission of potassium compared to that of rubidium. 

\begin{figure}
    \centering
    \includegraphics[width=0.5\linewidth]{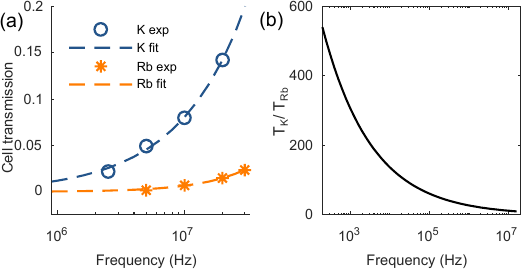}
    \caption{Calibrated RF electric field transmission curves, \textbf{(a)}, for both potassium (blue) and rubidium (orange) and \textbf{(b)} the ratio of the potassium and rubidium transmission curves, extrapolated beyond the few MHz range to qualitatively show the low frequency behavior. }
    \label{fig:Transmission}
\end{figure}

Beyond visualization, the applied field relative to the observed Stark shift can be measured to calibrate the field transmission at these higher frequencies. This calibration as a function of frequency is heuristically fitted to an exponential for both rubidium and potassium, shown in Figure~\ref{fig:Transmission}(a). In the range of the measurements, the potassium field transmission is already a factor of ten higher than that in rubidium. If the exponential trend is extrapolated to lower frequencies, as in Figure~\ref{fig:Transmission}(b), then potassium achieves a hundred times greater field transmission around 25~kHz. 

RF field sensitivity measurements show that increased transmission of potassium translates into improved sensitivity compared to rubidium. In these measurements, a weak signal with a frequency between 50~kHz and 10 MHz is interfered with a local oscillator with a slight offset frequency, in this case 10~kHz, creating a beat-note at the offset frequency. This beat-note is imprinted onto the optical transmission spectrum of Rydberg atoms, which can then be monitored as a function of the weak signal's amplitude~\cite{10.1063/1.5095633}. More details can be found in Section~\ref{se:Methods}. In this section, we specify the receiver sensitivity to an external field, whose value is obtained by dividing the applied AC voltage by the parallel plate separation, which provides an upper limit on the field that the vapor cell experiences. This will be discussed in more detail in Section \ref{sec:Discussion}. 

\begin{figure}
    \centering
    \includegraphics[width=0.5\linewidth]{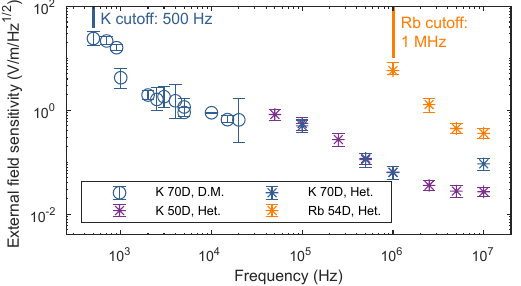}
    \caption{External field sensitivity measurements of both potassium and rubidium.}
    \label{fig:sensitivity}
\end{figure}

The sensitivity of heterodyne measurements at the 50D potassium state are shown as purple in Figure~\ref{fig:sensitivity}. In potassium, the measurements at 5~MHz and 10~MHz show very similar values, but then the sensitivity begins to drop, slowing near 200~kHz. The sensitivities of the rubidium measurements are shown as orange stars in Figure \ref{fig:sensitivity}. The sensitivity drops very quickly with frequency, and at 1~MHz, the function generator for the local oscillator maxes out at an external field amplitude of 300 V/m, preventing comparable sensitivities measurements at frequencies below 1~MHz. Between  1~MHz and 10~MHz, the sensitivity in rubidium is about an order of magnitude lower than that of potassium.  

\subsection{Potassium low-frequency sensing}
To further explore the limit of the low-frequency sensing capabilities of potassium, measurements are taken at the 70D state to increase the DC polarizability. Heterodyne measurements between 50 kHz and 10 MHz are shown as blue stars in Figure~\ref{fig:sensitivity}. For frequencies below 1 MHz, the heterodyne measurements show a marginal improvement in the sensitivity over the 50D state. However, at 10~MHz, the 70D shows a substantially reduced sensitivity. This arises because the Stark shifting is so strong that the spectral amplitude disperses over a large detuning, limiting the high-field sensitivity of the measurements. Moving below 50~kHz with heterodyne becomes challenging, as the signal frequency approaches the beatnote frequency. 

\begin{figure}
    \centering
    \includegraphics[width=0.5\linewidth]{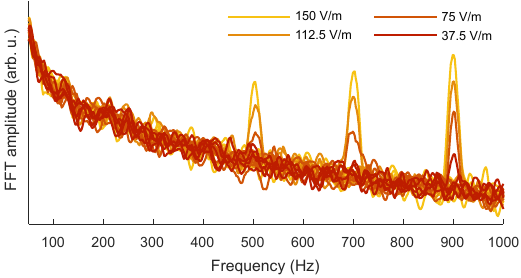}
    \caption{Plots of the Fourier transform beat note amplitudes of directly modulated EIT transmission at sub-kHz frequencies. }
    \label{fig:ULF}
\end{figure}

For measurements below 50~kHz, the RF signal is measured as a direct modulation (DM) of the EIT transmission at the frequency of the applied RF field. This is done by acquiring a time domain waveform and its Fourier transform and observing variations in the Fourier spectral power density as a function of the applied field. For potassium, we are able to measure DM sensitivities from 20~kHz down to 500~Hz, shown as blue circles in Figure~\ref{fig:sensitivity}. To exaemplify the behavior, the amplitude and frequency dependent Fourier components are shown in Figure~\ref{fig:ULF} for frequencies between 300~Hz and 900~Hz. These results typify the sub-50~kHz behavior. Namely, higher frequencies show both an increased spectral amplitude for a given field strength, as well as a reduced 1/f noise floor. This is captured in the final sensitivity values over this frequency range.

Measurements 300~Hz show a nonzero amplitude at high fields (yellow). However, the strong 1/f noise make it difficult to extract a meaningful sensitivity value.

\section{Methods}\label{se:Methods}
All of the experiments used identical beam paths and components with the sole exception of the coupling light, which could not be provided for both rubidium and potassium from a single source. To make the comparison as direct as possible, the two coupling sources are launched from the same fiber collimator, with all other components left unaltered. For potassium, the 456~nm light comes from an external cavity diode laser, and for rubidium, the 480~nm light comes from a cavity-based SHG diode laser. The probe light for both potassium and rubidium is sourced from the same external cavity diode laser. To generate an EIT signal, the coupling beam and counter propagating probe beams are spatially overlapped inside the vapor cells. The diameters of the probe and the coupling beam are 200~$\mu$m and 210~$\mu$m, respectively. 

\begin{figure}    
    \centering
    \includegraphics[width=0.5\linewidth]{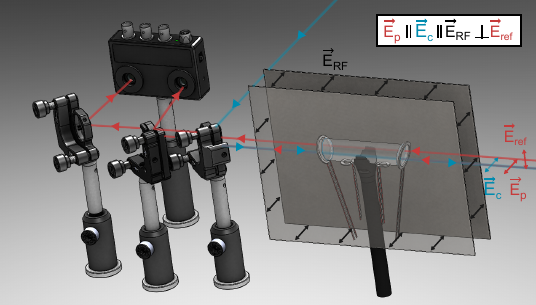}
    \caption{\textbf{(a)} Schematic of the experimental setup and the excitation scheme and energy levels for potassium \textbf{(b)} and rubidium \textbf{(c).} }
    \label{fig:exp}
\end{figure}

The most critical aspects of the experimental setup are shown in Figure~\ref{fig:exp}(a). The beams overlap inside a 75~mm fused silica vapor cell. RF fields are applied to a pair of 150~mm x 200~mm parallel plates that surround the vapor cell. The plate separation is 33~mm. The higher melting temperature of potassium requires heating to 60\textdegree~C to obtain an appreciable signal. Ceramic heaters and a 3D printed, non-conductive vapor cell mount reduce perturbations to the applied field. However, metallic wires are still necessary for delivering power to the heaters. Ceramic heating elements are also present in the rubidium experiments and it is slightly heated to 35\textdegree~C to achieve densities comparable to those of the potassium cell. Data are recorded simultaneously on a spectrum analyzer and oscilloscope for the respective experiments. 

An inverted ladder EIT scheme promotes electrons into their atomic Rydberg states. For potassium, shown in Figure~\ref{fig:exp}(b), the probe laser is locked to the 4S$_{1/2}$(F~=~2) state to the 4P$_{3/2}$ (F~=~3)  state, and the coupling laser is scanned across the 70/50D state, depending on the measurement. For rubidium, the probe is locked to the 5S$_{1/2}$(F~=~3) transition to the 5P$_{3/2}$ (F~=~4) state, and the coupling is scanned across the 54D state. 

Several spectroscopic adjustments are used to achieve as similar conditions as possible in the comparison between rubidium and potassium. The polarizabilities of the potassium and rubidium Rydberg states are calculated, and it is found that the 54D state of rubidium has a similar polarizability, 366~MHz/(V/cm)$^\textrm{2}$, to the 50D state of potassium, 361~MHz/(V/cm)$^\textrm{2}$. Choosing states of similar polarizability constrains origin of the vapor cell's low-frequency transmission to the atomic species. Furthermore, the Alkali Rydberg Calculator (ARC)\cite{sibalic_arc_2017} is used to calculate the transition dipole moments between the 4S$_{1/2}$ and 4P$_{3/2}$ state of potassium, 5.017~a$_0$e, and then 5S$_{1/2}$ to 5P$_{3/2}$ state of rubidium, 5.17~a$_0$e. The transition dipole moments for the coupling beams are also calculated. From potassium's 4P$_{3/2}$ to 50D state, the dipole matrix amplitude is 0.014~a$_0$e, and from rubidium's 5P$_{3/2}$ to 54D state, the dipole matrix element is found to be 0.022~a$_0$e. These calculation are used to normalize optical powers between the two species. As the probe dipole moments are close, the probe powers entering both rubidium and potassium vapor cells are set to 10~$\mu$W. Equalized Rabi rates for the coupling beams is achieved by using 120~mW of coupling power for the potassium experiment and 75~mW for rubidium. This results in Rabi rates of the probe and coupling beams of 44~MHz and 12~MHz respectively. 

For the RF heterodyne measurements, a 10~kHz beat frequency~\cite{10.1063/1.5095633} allows us to measure under identical conditions from 50~kHz to 10~MHz. Measurements using beatnotes of 20~kHz and 50~kHz produced marginally better sensitivities but introduced a larger gap between the DM and heterodyne measurements. The local oscillator (LO) amplitude is optimized by searching for the maximum beatnote amplitude as a function of LO power for each frequency. The peak position is optimized within 0.25~V. The signal amplitude is chosen as the minimum amplitude where a maximum in the beatnote versus LO amplitude scan is observed. The coupling laser is scanned over the EIT spectra while simultaneously measuring the modulation amplitude at the beat note frequency using a spectrum analyzer set to zero span. This bypassed any need for locking the coupling laser's frequency, which is not feasible with the limited power of the external cavity diode laser. The electric field amplitude is then placed into a logarithmic scale, and a linear fit of the beat note amplitude is acquired as a function of the RF amplitude. The intercept of this line with the noise floor defines the minimum field, and dividing the minimum field by the square root of the resolution bandwidth (RBW) of the measurement gives the final sensitivity. 

The DM measurements are more affected by our inability to lock the coupling laser. To handle this experimental limitation, the span of the coupling laser frequency scan is set to be a small fraction of the EIT line width, and a time-domain waveform is acquired. The time-domain waveform is Fourier transformed into the frequency domain, and the applied frequency peak amplitude is isolated and measured as a function of the RF amplitude and frequency. Similar to heterodyne measurements, the peak of the FFT amplitude is fit as a function of the log of the applied field amplitude, and the intercept with the noise floor gives the minimum field. As the measurements are performed in the time domain, multiplying by the square root of the acquisition time, effectively one over the RBW, gives the sensitivity.

For the measurements below 1~kHz, 128 spectra are sampled, and for all other measurements, 32 spectra are sampled at each frequency and amplitude. The error bars are found by extrapolating the a line found  by fitting the mean of the repeat scans $\pm$ one standard deviation (SD) of the repeat scan values at all field amplitudes. These SD offset amplitudes are fit as function of field, just as with the sensitivity mean data, and their intercept with the noise floor represents the error bars. This is beneficial as it considers how the deviation in the measurements at different amplitudes varies, which alters the slope intercept with the noise floor nonlinearly.  

\section{Discussion}\label{sec:Discussion}

\begin{figure}
    \centering
    \includegraphics[width=0.5\linewidth]{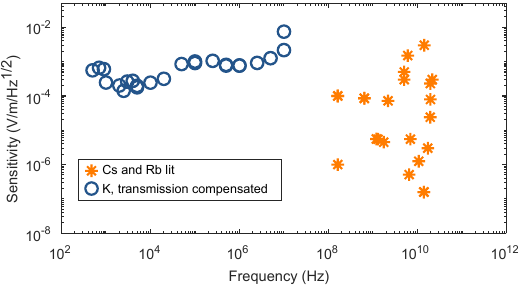}
    \caption{Transmission-compensated potassium EIT sensitivity measurements compared to literature sensitivities in rubidium and cesium. For equivalent comparison plated vapor cells have been excluded. Adapted from Ref.~\cite{schlossberger_rydberg_2024}.}
    \label{fig:Perspective}
\end{figure}

The difficulty of calibrating the field transmission between the electrodes and the vapor cell at low frequency means that we cannot measure the exact transmission at low frequencies. However, we can extrapolate the transmission curve at~MHz levels to the~kHz regime to gain insight into how these results compare with the literature. This is done in Figure \ref{fig:Perspective}. Although some uncertainty remains in the exact sensitivity values, it is clear that potassium Rydberg sensors enable a simple method for performing low-frequency sensing that has been left largely unexplored by the community. The absence of sensitivity measurements in rubidium and cesium highlights how the screening effect strongly limits rubidium and cesium vapor cells at sub-MHz frequencies. Although the demonstrated/projected sensitivities are higher compared to their heavier counterparts, these relatively simple measurements are not outside the range of reported sensitivities~\cite{schlossberger_rydberg_2024}, and could likely be improved with known methods from the field~\cite{brown_very-high-_2023, cai_sensitivity_2023, dixon_rydberg-atom-based_2023, holloway_rydberg_2022, kumar_atom-based_2017, prajapati_enhancement_2021, prajapati_high_2024}. Even in this light, these measurements represent a notable expansion in the capabilities of these quantum sensors. 

Although these measurements represent a substantial improvement over rubidium, it is also important to note that the absolute transmission is still rather poor, with 1\% passing through the vapor cell around 800~kHz. So, while these results do open up a highly accessible path for the community to investigate low frequencies, the problem of low-frequency transmission in vapor cells still requires considerable attention. 

Understanding the exact mechanism of why potassium has improved transmission at these low frequencies is beyond the scope of a single paper. With that said, we currently have a working hypothesis on the mechanism behind the screening. We hypothesize that the origin of the low-frequency field transmission of potassium arises from variations in the chemical interactions between the different alkali and the bulk of the vapor cell. The smaller atomic radius of potassium compared to rubidium and cesium would make potassium more likely to permeate into and through the glass~\cite{gershinskii_investigation_1980, samuneva_interaction_1989, cormack_alkali_2002, karlsson_trends_2017}. Despite this, the permeation of potassium ions into the glass does not increase the conductivity of the glass as much compared to the permeation of heavier alkali, supported not just by this work but also by other studies~\cite{richardson_extraction_2025}. Previous results have shown that cesium reacts strongly with non-bridging oxygen defects in silicon dioxide lattices~\cite{charles_polarization_1961, lopez_adsorption_1999}. In this reaction, the cesium atom donates almost a full electron charge to the surrounding lattice, which is considerably more than what a lighter alkali would give~\cite{lopez_adsorption_1999}. The full donation of the charges to the lattice paired with reduced permeability produces a dense volume of low-mobility charges, which results in a pronounced screening effect. For the lighter alkali, the higher permeability paired with a reduced charge donation to the lattice results in a more diffuse charge distribution in the lattice, effectively reducing the screening. This chemical permeation hypothesis is further supported by the previously mentioned works on the consumption of alkali in vapor cells~\cite{ma_modification_2009, woetzel_lifetime_2013}.

Many have proposed condensed alkali metal on the interior walls of the vapor cell as the primary mechanism behind the screening~\cite{jau_vapor-cell-based_2020, lim_kilohertz-range_2023, ma_study_2025}. However, in this picture, the volume of alkali inside the vapor cell would be a parameter that can be adjusted to control the low-frequency transmission. Furthermore, reports of reduced optical absorption with time in vapor cells~\cite{ma_modification_2009, woetzel_lifetime_2013} would then result in an improved low frequency field transmission as the number of atoms available for surface adsorption decreases. Additionally, if surface aggregation is the main factor contributing to the screening effects, potassium would exhibit more screening, since its electrical resistivity is 72~n$\Omega \cdot$m, compared to the resistivities of rubidium and cesium of 128 and 205~n$\Omega\cdot$m, respectively\cite{chi_electrical_1979}. These bulk values should be applicable at the alkali thicknesses predicted in vapor cells~\cite{fuchs_conductivity_1938, ma_modification_2009}. So, while these layers likely contribute to the screening effects, there are inconsistencies with these interpretations and the alkali properties that suggest an alternative mechanism is at play.

A deeper understanding of the mechanisms at play in the interactions of these vapor cells will likely require methods outside electrometry. The element selectivity of x-ray techniques~\cite{mastelaro_x-ray_2018} and UV-absorption~\cite{lopez_adsorption_1999, samuneva_interaction_1989} edges might also help quantify defect behavior before and after exposure to alkali, although the high-photon energy of these probes might induce defects and otherwise alter the lattice/glass structure meaningfully. NMR is another promising method that has uncovered many details in glasses~\cite{lopez_adsorption_1999, youngman_nmr_2018} with less damaging radiation. A further challenge to understanding these mechanisms is the innate anisotropy of glasses, whose traits can vary strongly with subtle changes in composition. This is further exacerbated as exposure of the glass to chemicals, UV light, heat, and strong electric fields substantially alter the glass behavior. Understanding the glass behavior under these conditions will be especially critical for wafer-level production of Rydberg sensors.

\section{Conclusion}\label{sec13}
These measurements have demonstrated electric field sensitivities at sub-kHz frequencies of an all dielectric, potassium Rydberg sensor. While much work is needed to improve the sensitivity of these devices, these results open a highly accessible avenue for the community to explore low-frequency sensing. Further, these results point to a need for a deeper understanding of the chemical interactions between alkali metals and vapor cells in the context of Rydberg sensors, where optimal performance might arise under different configurations compared to atomic clocks or optically pumped magnetometers. 

\backmatter

\bmhead{Supplementary information}



\bmhead{Author contributions}
D.H. contributed to the conceptualization, methodology, data acquisition, data analysis, figure generation, and paper writing. R.T., D.M., contributed to the data acquisition and methodology. E.M., M.H., N.S., N.P., M.S., S.B., A.A-G. contributed to the methodology. C.L.H contributed to the methodology and paper writing. 

\bmhead{Acknowledgments} This research  was supported by NIST under the NIST-on-a-Chip program. 

\bmhead{Competing interests} The authors declare no conflicts of interest.   A contribution of the U.S. government, this work is not subject to copyright in the United States. Any collaboration associated with the work does not imply recommendation or endorsement of any product or service by NIST, nor is it intended to imply that any such products or services are necessarily the best available for the purpose.

\begin{appendices}

\end{appendices}


\bibliography{PotassiumULF}

\end{document}